\documentclass[aps,prd,eqsecnum,twocolumn,epsf,nofootinbib,preprintnumbers]{revtex4}
\usepackage{amsfonts,latexsym,eucal,amsmath,amssymb,times,mathrsfs}
\usepackage[dvips]{graphicx}
\usepackage[pdf]{pstricks}

\begin{document}
\thispagestyle{empty}

\title{Cosmological imprints of string axions in plateau}
\author{Jiro Soda$^{1}$}
\email{jiro_at_phys.sci.kobe-u.ac.jp}
\author{Yuko Urakawa$^{2, 3}$}
\email{urakawa.yuko_at_h.mbox.nagoya-u.ac.jp}
\address{\,\\ \,\\
$^{1}$ Department of Physics, Kobe university, Kobe 657-8501, Japan\\
$^{2}$ Department of Physics and Astrophysics, Nagoya University, Chikusa,
Nagoya 464-8602, Japan\\
$^{3}$  Institut de Ciencies del Cosmos,
Universitat de Barcelona, Marti i Franques 1, 08028 Barcelona, Spain}

\preprint{KOBE-COSMO-17-13}


\begin{abstract}
We initiate a study on various cosmological imprints of string axions whose scalar potentials
 have plateau regions. In such cases, we show that a delayed
 onset of oscillation rather generically leads to a parametric resonance
 instability. In particular, for ultralight axions, the parametric resonance can enhance
 the power spectrum slightly below the Jeans scale, alleviating the tension
 with the Lyman $\alpha$ forest observations. We also argue that a
 sustainable resonance can lead to an emission of gravitational waves at the frequency bands which are detectable by gravitational wave interferometers
 and pulsar timing arrays and also to a succeeding oscillon formation.
\end{abstract}


\maketitle
{\it Introduction.}--
Compactifications in string theory generically predict various axions 
in 4D low energy effective field theory. Exploring  imprints of
axions in cosmology provides an important tool to probe
extra dimensions predicted in string theory~\cite{StAx}. 
Phenomenological impacts of axions have been
mostly studied by considering the quadratic potential. However, once
an axion is away from the potential minimum, the potential deviates
from the quadratic form. In particular, when the dilute instanton gas
approximation does not hold, the scalar potential can be more
flatten than the conventional cosine form~\cite{Dubovsky:2011tu,Nomura:2017ehb}. Therefore, it is worth investigating phenomenological consequences 
of axions with such plateau regions in their scalar potentials.

Distinctively, the axions which were located at such plateau regions
generically undergo a parametric resonance after their onsets of oscillations, which exponentially enhances
the modes in the resonance bands and potentially leaves various phenomenological
impacts. This parametric resonance instability has been widely studied in the
context of reheating after inflation. (For reheating, see e.g., historical papers \cite{preheating} and \cite{KLS}.) In Refs.~\cite{oscillon}, it was shown that when
the potential is shallower than the quadratic form, the instability leads to a
fragmented configuration of the oscillating scalar field, the so
called {\it oscillon} (see also Refs.~\cite{Antusch:2016con, Antusch:2017flz}). In
Ref.~\cite{Yoshida:2017cjl}, it was shown that the oscillating axion can 
induce resonance phenomena also in gravitational waves.

In string theory, there appear axions in a wide mass range. For example, the
large volume scenario predicts the presence of extremely light
axions (see e.g., Refs.\cite{LVS}). The onset time of the
oscillation varies, depending on the mass scale of the axion. In particular,
the ultra-light axion (ULA) whose mass is of $O(10^{-22} {\rm eV})$
commences the oscillation before the matter-radiation equality and behaves as a fuzzy dark
matter. The ULA has been often discussed in the context of the small
scale issues of $\Lambda$CDM~\cite{review}. Meanwhile, it was argued that the ULA with 
$m \simeq 10^{-22}$eV is marginally incompatible with the Lyman $\alpha$ forest
observations, since the ULA smooths out the small scale structures too
much~\cite{Irsic:2017yje}. It is interesting to see whether the
parametric resonance instability can relax the tension with the Lynman $\alpha$
forest observations or not.

{\it Setup of problem.}-- In order to study dynamics of string axions whose potentials have a shallower region
than the quadratic potential, we consider a canonical scalar field
$\phi$ with a scalar potential $V(\phi)$ given by
$V(\phi) = (m f)^2\, \tilde{V}( \tilde{\phi})$ with $\tilde{\phi} \equiv \phi/f$. 
Here, $\tilde{V}(\tilde{\phi})$ satisfies the following properties: i)
$\tilde{V}(\tilde{\phi}) \to \tilde{\phi}^2/2$ in the limit
$\tilde{\phi} \to 0$, ii) $\tilde{V}(\phi)/\tilde{\phi}^2 \to 0$ in the
limit $|\tilde{\phi}| \to \infty$. Since the axion is a pseudo-scalar, it is
reasonable to impose additionally $Z_2$ symmetry on the potential. The parameter $f$ agrees with the
decay constant in the case with the cosine potential. Roughly speaking,
$m$ determines the onset time of the oscillation (under a certain
initial condition) and $f$ determines the energy density for a given
$m$.

In this paper, we will
investigate phenomenological consequences of the axions
with a potential which satisfies the conditions i) and ii). Considering a situation where the self-interaction is much more important than the gravitational interaction, we solve the Klein-Gordon (KG)
equation 
\begin{align}
 & \Box \phi - V_{\phi}=0
 \end{align}
in a fixed geometry.

{\it Evolution of the homogeneous mode.}-- First, we consider the
evolution of the homogeneous mode of $\phi$. In the following, we assume
the background expansion law as $a \propto t^p$ with $p> 0$~\footnote{When the axion dominates the universe, we need to determine the expansion law consistently. However, this point is not very important in discussing the dynamics of inhomogeneous modes, especially when the onset of the oscillation is delayed.}. When
the field  $\phi$  does not dominate the universe, the power $p$ cannot be
determined only from the dynamics of $\phi$. Then, the
KG equation for the homogeneous mode is given by 
\begin{align}
 & \frac{d^2 \tilde{\phi}}{d x^2} + \frac{3 p}{x} \frac{d
 \tilde{\phi}}{dx} + p^2 \frac{d \tilde{V}}{d \tilde{\phi}} = 0  \label{Eq:hom}
\end{align}
with $x \equiv m/H = m t/p$. Notice that all the dimensionful parameters
dropped out from the equation and the onset time of oscillation, $x_{osc}$, is solely
determined by the initial conditions $\tilde{\phi}_i \equiv \tilde{\phi}(x_i)$
and $\tilde{\phi}_{x,i} \equiv d \tilde{\phi}/dx (x_i)$. The Hubble
parameter at $x_{osc}$ for a given $m$ is determined as $H_{osc} = m/x_{osc}$.
When the axion stays at a plateau, it behaves as a
cosmological constant. In fact, when the potential gradient term
is negligible, the equation of motion (\ref{Eq:hom})
can be solved analytically as 
$\tilde{\phi}(x) = C_1 + C_2\, x^{1-3p}$. Meanwhile, around the bottom
of the potential with $|\tilde{\phi}| \ll 1$, Eq.~(\ref{Eq:hom}) can be
also solved analytically. However, in general, Eq.~(\ref{Eq:hom}) can be
solved only numerically in the intermediate range.

\begin{figure}[tbh]
\begin{center}
\begin{tabular}{cc}
\includegraphics[width=7cm]{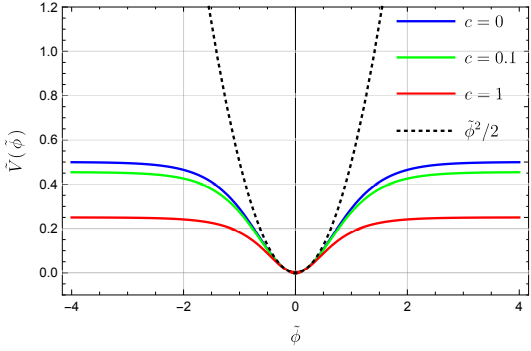}
\end{tabular}
\caption{The plot shows the potential shape of the $\alpha$ attractor
 type potential (\ref{Exp:V}) for different values of $c$.}
\label{Fg:Vtanhs}
\end{center}
\end{figure}
\begin{figure}[tbh]
\begin{center}
\begin{tabular}{cc}
\includegraphics[width=7cm]{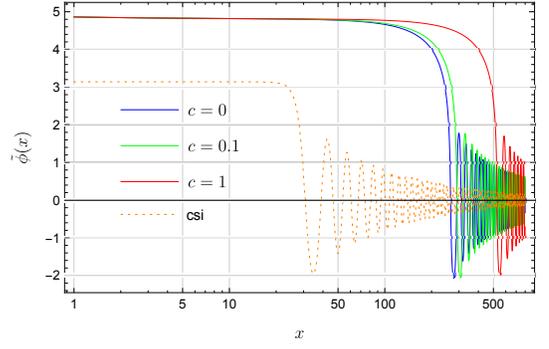}
\end{tabular}
\caption{This plot shows the time evolution of $\tilde{\phi}$ in RD.}
\label{Fg:phitans}
\end{center}
\end{figure}

Notice that when $\phi$ is located at the plateau region at the onset of oscillation, the oscillation does not necessarily take place around $x \simeq 1$ or 
$H \simeq m$. What will be discussed in this paper applies rather
generically, in case the scalar potential satisfies the conditions i)
and ii). However, for a concrete analysis, as an example, we consider an $\alpha$
attractor type potential, given by 
\begin{align}
 & \tilde{V}(\tilde{\phi}) = [1 + c
 (\tanh \tilde{\phi})^{2}]^{-1} \times  (\tanh \tilde{\phi})^2/2 \label{Exp:V}
\end{align}
with $c \geq  0$. The $\alpha$ attractor model was
considered as a generalization of the superconformal
models~\cite{alphaatt}. This potential is shown in Figure
\ref{Fg:Vtanhs} for different values of $c$.  For $|\tilde{\phi} | < 1$, the second derivative of the potential is given by
$ \tilde{V}_{\tilde{\phi} \tilde{\phi}} = 1 - 2(2 + 3 c) \tilde{\phi}^2 + O(\tilde{\phi}^4)$, i.e., 
the curvature of the potential becomes smaller for a larger value. Figure \ref{Fg:phitans} shows that the oscillation starts at $x_{osc} \gg 1$, when
we choose the initial condition $\tilde{\phi}_i=5$ and
$\tilde{\phi}_{x,i}=-1$, starting at the plateau region. As was later pointed out in Ref.~\cite{KSU}, $x_{osc}$ can be roughly estimated as $x_{osc} \sim \sqrt{|\tilde{\phi}_i/ \tilde{V}_{\tilde{\phi}} (\tilde{\phi}_i )|}$,  indicating that the onset of the oscillation delays, taking $x_{osc} \gg 1$, when $\phi$ was initially located at a potential region which is much shallower than the quadratic potential. (When the
plateau is wide enough, the evolution does not much depend on
the initial velocity because of the over damping.) The orange
dotted line shows the time evolution for the conventional
cosine potential $\tilde{V}(\tilde{\phi}) = 1 - \cos \tilde{\phi}$ with $\tilde{\phi}_i= \pi$ and $\tilde{\phi}_{x,i}=
- 10^{-4}$. Even with this fine-tuned initial condition, the oscillation starts
much earlier than the plateau case.

When $\phi$ starts to oscillate before the matter-radiation equality, we can estimate the decay constant by 
equating the energy density of the radiation
$\rho_\gamma^{eq}$ with that of dark matter 
$\rho_m^{eq} =\rho_\phi^{eq}/\beta_\phi$, where $\beta_\phi$ denotes the
fraction of the ULA among the total dark matter, as
\begin{eqnarray}
  f \simeq \frac{\beta_\phi^{\frac{1}{2}}}{m} \left( \frac{\rho_{eq}^{\frac{1}{3}}
 H_{osc}^2 }{8 \pi G } \right)^{\!\frac{3}{8}}\!\! \simeq \left(
 \frac{10^{-22}{\rm eV}}{m} \right)^{\frac{1}{4}} \frac{\beta_\phi^{\frac{1}{2}}}{x_{osc}^{\frac{3}{4}}}\, 10^{17} {\rm GeV}  \,,  
 \label{decay}
\end{eqnarray}
where the quantities with the index $eq$ denote those at the equal time
and the quantities with the index $osc$ denote those at the onset of the
oscillation. Here, taking into account that the kinetic energy and the potential
energy are comparable at the onset of the oscillation, we used 
$\rho_{\phi}^{osc} \simeq 2 V_{osc} \simeq (m f)^2$. Thus, once
$x_{osc}$ is given by solving Eq.~(\ref{Eq:hom}), the decay constant $f$ is determined by Eq.~(\ref{decay}) for a given $m$.

{\it Parametric resonance instability.}-- Next, we study the evolution of
the inhomogeneous modes. The perturbed KG equation
for the axion is given by
$\ddot{\varphi} +  3H \dot{\varphi} + (k/a)^2  \varphi +  V_{\phi \phi} \varphi =0$,
where $\varphi$ is the perturbed axion field and we neglected the metric perturbations. According to our numerical
analysis, the metric perturbation does not play a crucial role at least in the
early stage of the resonance instability, where the linear analysis can
apply. Again, we can rewrite the perturbed KG equation for
$\tilde{\varphi} \equiv \varphi/\varphi_i$, which is normalized by the
initial value $\varphi_i$, in a dimensionless form 
\begin{align}
 & \frac{d^2  \tilde{\varphi}_k}{dx^2} + 3\, \frac{p}{x} \frac{d
 \tilde{\varphi}_k}{d x} + p^2  \tilde{k}^2 \left( \frac{x_i}{x}
 \right)^{2p} \tilde{\varphi}_k + p^2 \tilde{V}_{\tilde{\phi}
 \tilde{\phi}}\, \tilde{\varphi}_k = 0\,, \label{Eq:mode}
\end{align}
where we used $k/(am) = \tilde{k}\, (x_i/x)^p$ with 
$\tilde{k} \equiv k/(a_i m)$. Depending on the choice of the initial
time, the corresponding wavenumber $\tilde{k}$ varies, while $k/a$ is
independent of the choice of the initial time. Here, we choose $x_i=1/10$.

\begin{figure}[tbh]
\begin{center}
\begin{tabular}{cc}
\includegraphics[width=7cm]{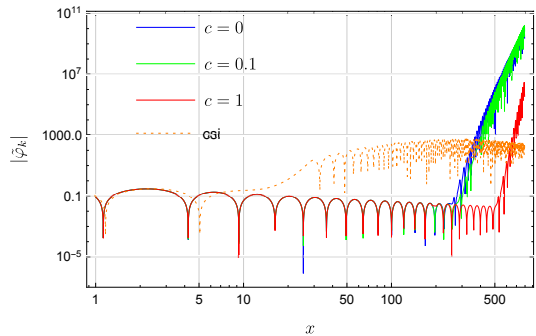}
\end{tabular}
\caption{
This plot shows the time evolution of $\tilde{\varphi}_k$ with
 $\tilde{k}=10$ under the initial condition $\tilde{\phi}_i=5$ during RD.}
\label{Fg:dphitansvc}
\end{center}
\end{figure}
\begin{figure}
\begin{center}
\begin{tabular}{cc}
\includegraphics[width=7cm]{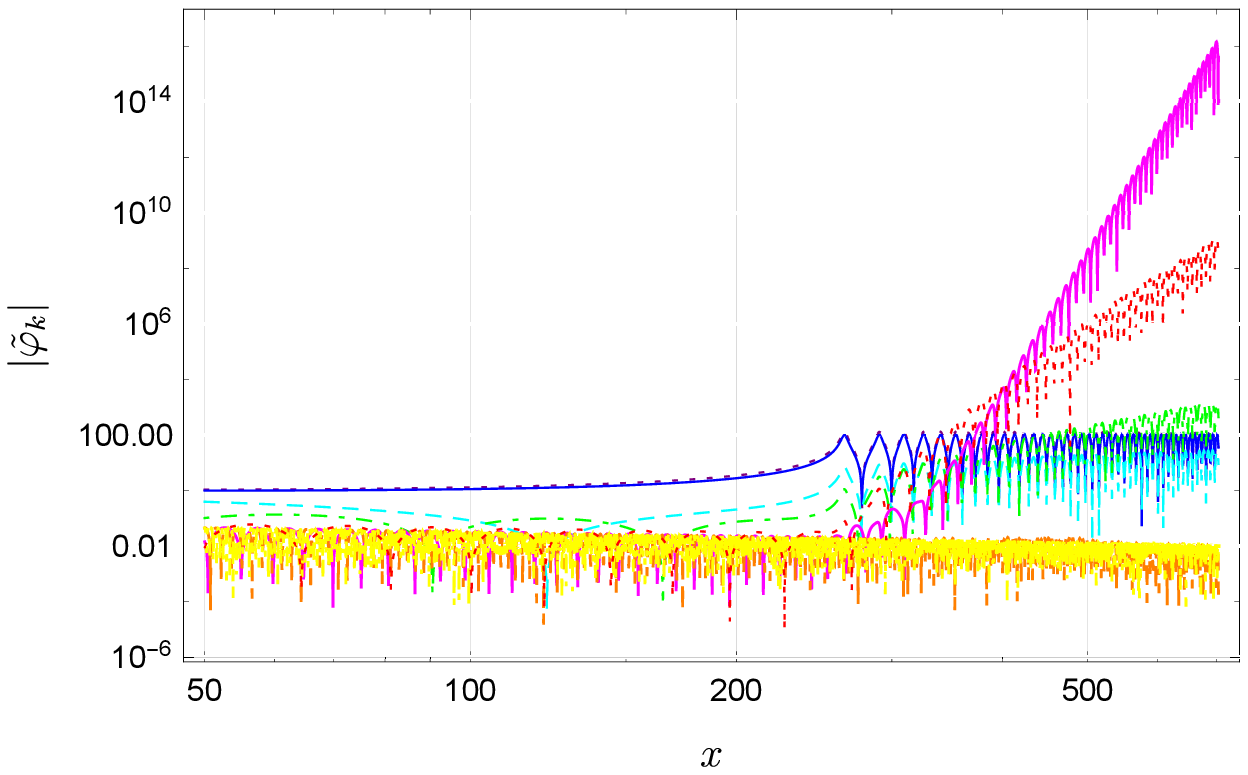}\\
\includegraphics[width=7cm]{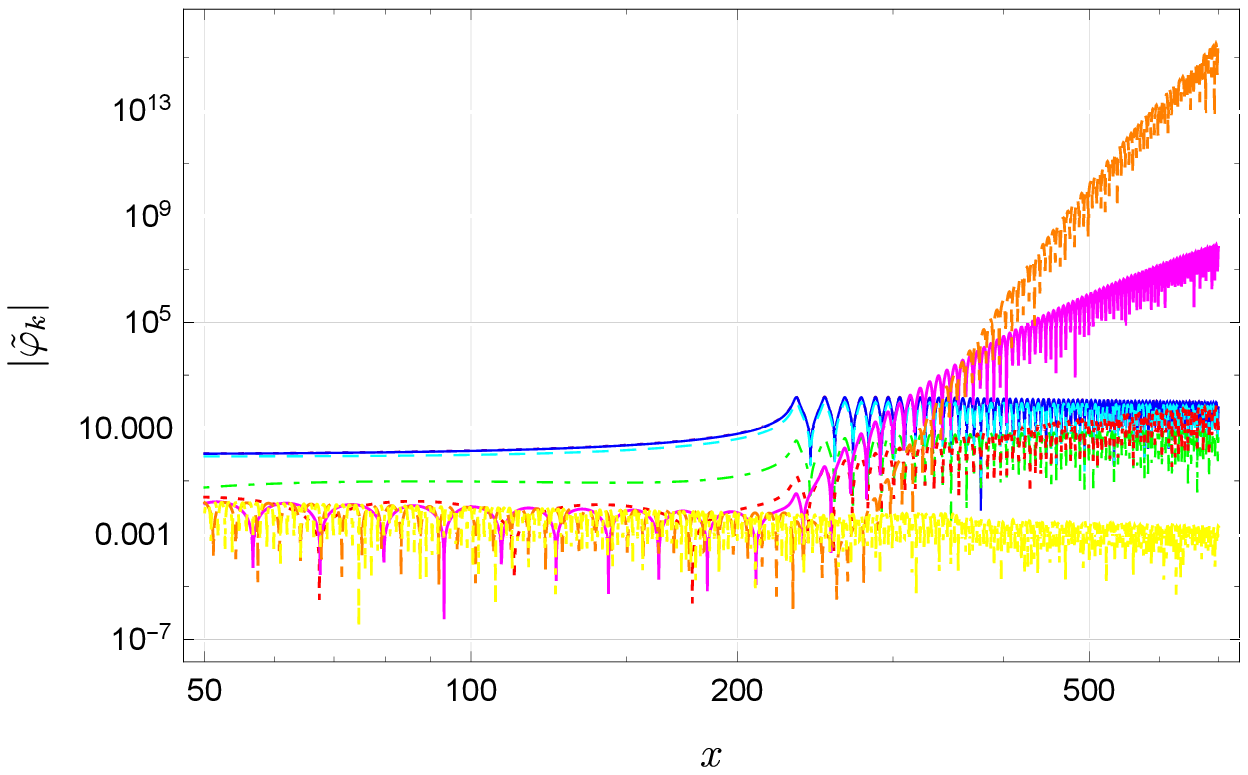}
\end{tabular}
\caption{
These plots show the time evolutions of $\tilde{\varphi}_k$ during RD
 (up) and MD (bottom) for different wavenumbers
 $\tilde{k}$: $\tilde{k}=10^{-1}$ (Purple, Dotted), $\tilde{k}=10^{-1/2}$
 (Blue), $\tilde{k}=1$ (Light blue, Dashed), $\tilde{k}=10^{1/2}$
 (Green, Dotdashed),
 $\tilde{k}=10$ (Red, Thick-dotted), $\tilde{k}=10^{3/2}$ (Magenta, Thick), $\tilde{k}=10^2$
 (Orange, Thick-dashed), and  $\tilde{k}=10^{5/2}$ (Yellow, Thick-Dotdashed). Here, we choose $c=0$. The parametric resonance instability is slightly more efficient in RD than in MD, since the comic expansion is slower. }
\label{Fg:dphitans}
\end{center}
\end{figure}

Figure \ref{Fg:dphitansvc} shows the time evolution of
$\tilde{\varphi}_k$ for the same potentials as in Fig.~\ref{Fg:phitans} during RD. The fluctuation $\tilde{\varphi}_k$ for the cosine potential under the same initial condition as in Fig.~\ref{Fg:phitans} grows much less than the one for the $\alpha$ attractor type potential.  Figure \ref{Fg:dphitans} shows the time evolution of $\tilde{\varphi}_k$ for different wavenumbers
$\tilde{k}$ during RD (up) and MD (bottom), when $\tilde{V}$ is given by
Eq.~(\ref{Exp:V}) with $c=0$. The modes with $\tilde{k}=10^{-1}$ and
$\tilde{k}=10^{-1/2}$ got slightly enhanced just after the onset of the
oscillation due to the tachyonic instability. However, the conventional tachyonic instability is not very
efficient, because the second derivative of the potential soon starts to
oscillate, taking both positive and negative values. In our accompanying
paper \cite{KSU},  where we performed a more detailed analysis, we
showed that a different type of resonance instability, dubbed the {\it
flapping resonance instability} sets in, when the curvature of the
potential oscillates between negative and positive values~\footnote{For
the $\alpha$ attractor potential with larger values of $c$, the flapping
resonance tends to be more prominent than the narrow
resonance~\cite{KSU}.}.

In order to understand the instability more intuitively, here let us
analyze the equation (\ref{Eq:mode}), neglecting the Hubble
friction. Around the bottom of the potential, i.e., 
$\tilde{\phi} < 1$, the homogeneous mode of the axion oscillates with the frequency 
$|\tilde{V}_{\tilde{\phi} \tilde{\phi}}|^{1/2} \simeq 1$ and the
solution is given by $\tilde{\phi}= \tilde{\phi}_* \cos z$ with $z=mt$. Using this solution, Eq.~(\ref{Eq:mode}) is given by Mathieu equation as
\begin{align}
 & \frac{d^2}{ d z^2} \tilde{\varphi} + (A - 2 q \cos 2z) \tilde{\varphi} =0\,,
\end{align} 
where $A$ and $q$ are defined as
\begin{align}
 & A \equiv \left( \frac{k}{m\, a_{osc}}
 \right)^2\!\! + 1 - (2 + 3c) \tilde{\phi}_*^2 , \quad
 q \equiv \frac{2 + 3 c}{2} \, \tilde{\phi}_*^2. \nonumber 
\end{align}
Here, $a_{osc}$ denotes the scale factor at around the onset of the
oscillation. The parametric resonance takes place for the narrow band $A \simeq n^2$, where $n$ is
an integer. The width of the resonance band is proportional to
$(q/A)^n$. The dominant growing mode is the $n=1$ mode and the growth
rate $\gamma$ with $\tilde{\varphi} \propto e^{\gamma x}$ is given by
$\gamma \simeq q/2 = (2 + 3c) \tilde{\phi}_*^2/4$. Notice also that the resonance
band becomes wider for a larger $c$ as shown in Fig.~\ref{Fg:dphitansvc}.
The resonance wavenumbers can be predicted from the first resonance band of
Mathieu equation, leaving aside factor deviations.

The cosmic expansion makes the parametric
resonance instability inefficient mainly due to the following two
effects: first, the amplitude $\tilde{\phi}_*$ decreases due to the Hubble
friction, reducing the growth rate and second, more importantly here,
the physical wavenumbers in the resonance band(s) are red shifted. When the gradient of the potential is
shallower, the onset of the oscillation gets more delayed, i.e.,
$x_{osc} \gg 1$. Then, when the parametric resonance instability sets
in, the redshift of the resonant modes due to the cosmic expansion is
not effectively important any more. This leads to the 
sustainable resonant growth without being disturbed by the cosmic
expansion. Because of that, an efficient resonance instability
requires a shallower potential than the quadratic potential region, where the
oscillation takes place at $x_{osc} \simeq 1$. The exponential
growth continues until the time when the re-scattering due to the
backreaction becomes important, i.e., $|\varphi/f| \simeq O(1)$
(see, e.g., Ref.~\cite{KLS}). As is shown in Fig.~\ref{Fg:dphitansvc}, for a
larger $c$, the resonance instability proceeds more rapidly, since
the oscillation starts later. To visualize this aspect more clearly, for a reference, we also plotted the time evolution of
$\tilde{\phi}$ and $\tilde{\varphi}$ for the cosine potential in
Fig.~\ref{Fg:phitans} and Fig.~\ref{Fg:dphitansvc}, respectively.

In Fig.~\ref{Fg:dphitansvc} and Fig.~\ref{Fg:dphitans}, where we choose the
initial field value $\tilde{\phi}_i = 5$, the parametric resonance
persists without being disturbed by the cosmic expansion. On the
other hand, when we choose a smaller value of $|\tilde{\phi}_i|$ as an
initial condition, the parametric resonance can persist only in a
shorter period, leaving only a milder enhancement of the fluctuation.

{\it Jeans scale.}-- 
It is known that the ULA has an emergent pressure on small scales
and the Jeans wavenumber is given by
$
k_J (a) \simeq \sqrt{m H}\, a\,,   
$
where we used $c_s \simeq  k/(2 m a)$~\cite{review}.
When the ULA dark matter becomes the dominant component of the universe for $a > a_{eq}$, the
structures below the Jeans length are smoothed out. As is shown in Fig.~\ref{Fg:dphitans}, the parametric
resonance instability takes place for the wavenumbers slightly
above the Jeans wavenumber. This can be understood as
follows. The resonance wavenumber in the first band $k_r$ satisfies 
$k_r/(m a_{osc}) \simeq O(1)$. Therefore, using 
$k_J/(m a) \simeq 1/\sqrt{x}$, we find a universal relation $k_r \simeq \sqrt{x_{osc}}\, k_J(a_{eq})$.   
When the scalar potential of the ULA dark matter has a plateau region,
the parametric resonance which takes place for the smaller scales than
the Jeans scale can enhance the perturbation of the ULA before the
matter-radiation equality. When $x_{osc}$ is not too large, after
the matter-radiation equality, $k_r$ soon becomes smaller than $k_J$,
which increases as $a^{1/4}$ in MD. Then, the density perturbation of
the ULA with $k_r$ starts to grow due to the Jeans instability. 
Therefore, a mild enhancement also can supply the missing small scale
structures, asserted in Ref.~\cite{Irsic:2017yje}. In fact, in
Ref.~\cite{Schive:2017biq}, it was argued that a mild enhancement 
around $k_J$ can lead to a significant enhancement of the low-mass halo abundance
even for the conventional cosine potential by accepting a careful tuning
of the initial condition~\cite{cos}. By contrast, in the plateau case, we can evade the
fine-tuning issue.

{\it GWs emission and Oscillon formation.}-- When the resonance
instability continues, the linear perturbation ceases to be a good
approximation, even if we start with an almost homogeneous initial
condition. In the subsequent and transient stage, the oscillating axion in a highly inhomogeneous spatial configuration leads to a prominent emission of the gravitational waves
(GWs)~\cite{Zhou:2013tsa, Antusch:2016con,Antusch:2017flz}. In contrast to the
GWs emitted during the reheating, the
peak frequency of the GW spectrum emitted later times
can be in sensitivity bands of GW detectors. Here, we roughly evaluate the peak
frequency of the GW emitted either during RD or MD (the later MD) as
$f_0 \simeq m/(1+z_*)$, where $z_*$ denotes the redshift at the
emission. In the following, for simplicity, we identify the Hubble parameter at the emission
as the one at the onset of the oscillation, i.e., 
$H(z_*) \simeq  m/x_{osc}$, assuming that the GW emission takes
place immediately after $x=x_{osc}$ in the cosmological time scale. Then, the frequency of the GWs emitted during RD
can be given by
\begin{align}
 & f_0 \simeq \left( \frac{m}{10^4 {\rm eV}}\, x_{osc}
 \right)^{\frac{1}{2}} {\rm Hz} \qquad ({\rm RD})\,,
\end{align}
where we used $(1+z_*) \simeq (H(z_*)/H_{eq})^{1/2} (1 + z_{eq})$. Similarly, using 
$(1+z_*) \simeq (H(z_*)/H_{eq})^{2/3} (1 + z_{eq})$, we obtain the
frequency of the GWs emitted during MD as 
\begin{align}
 & f_0 \simeq \left( \frac{m}{10^{20} {\rm eV}}\, x^2_{osc}
 \right)^{\frac{1}{3}} {\rm Hz} \qquad ({\rm MD})\,.
\end{align}
When we avoid choosing a hierarchically large value of $|\tilde{\phi}_i|$, $x_{osc}$ ranges in
$O(1) \alt x_{osc} \alt O(10^4)$. For a direct detection, the GWs emitted during RD is more promising, e.g., for 
$m \simeq 10^{-6}$eV and $x_{osc} \simeq 10^4$, the frequency $f_0$ is
in the band of space interferometers~\cite{GWspace} 
and for $m \simeq 10^3$eV and $x_{osc} \simeq 10^4$, $f_0$ is in the
band of ground based interferometers~\cite{GWground}. (When axions have unsuppressed
interactions with the electromagnetic field, the larger mass range should be excluded because
of the photon decay process~\cite{StAx, review}.) Meanwhile, for $m \simeq 10^{-16}$eV, 
$f_0$ is in the detectable range by pulsar timing
arrays~\cite{PTA}.

In order to compute the amplitude of GWs, we
introduce a parameter $\epsilon ( \leq 1)$ which denotes the ratio between the
relative spectral energy density of the emitted GWs and that of the homogeneous axion at
the onset of the oscillation, which is of $O((mf)^2)$. Assuming that GWs
were emitted just after the onset of the oscillation (in cosmological
time scales), we obtain $\Omega_{gw}$ as~\cite{Khlebnikov:1997di}
\begin{eqnarray}
  \Omega_{gw} \simeq \epsilon \Omega_r \frac{(mf)^2 }{\rho_{eq}} \!\left(\frac{a_{osc}}{a_{eq}}\right)^4\!
  \simeq 10^{-8}\, \epsilon x_{osc}^2\! \left( \frac{f}{10^{16} {\rm GeV}} \right)^{\!\!2}\!.  
\end{eqnarray}
In particular, for $m \agt 10^{-27}$eV, the axion starts to oscillate
before the equal time, behaving as a dark matter component. Then,
using Eq.(\ref{decay}), we obtain
\begin{eqnarray}
  \Omega_{gw} 
  \simeq 10^{-4} \epsilon \beta_\phi \sqrt{\frac{x_{osc}}{10^4}} \, \sqrt{\frac{10^{-22}{\rm eV}}{m}}\,.
\end{eqnarray}
When the axion is the dominant component of dark matter, i.e.,
$\beta_\phi \simeq 1$ and $x_{osc}= O(10^4)$, we obtain 
$\Omega_{gw} \simeq  10^{-7} \epsilon$ for $m \simeq 10^{-16}$eV. Meanwhile, we
obtain $\Omega_{gw} \simeq 10^{-12} \epsilon$ for $m \simeq 10^{-6}$eV  
and $\Omega_{gw} \simeq 10^{-16} \epsilon$ for $m \simeq 10^3$eV. Therefore,
for $\epsilon \agt 10^{-6}$, we can expect a detection of GWs
emitted by the resonantly oscillating modes of the string axions, using the pulsar timing
arrays.

In Ref.~\cite{KSU}, to evaluate the GW amplitude accurately, we
have conducted the lattice simulation. Our analysis indicates that axions which were initially located at plateau regions potentially lead to a detectable emission of GWs, opening a new window
of string axiverse.

Meanwhile, in Refs.~\cite{oscillon}, it was numerically
shown that the long-lasting instability can lead to a fragmented configuration of the
oscillating scalar field. Since the cosmic
expansion is not crucial at the oscillon formation, clumps of the 
oscillon can be formed also during RD and the later MD. This will be also shown in Ref.~\cite{KSU}.

{\it Summary: New window in string axiverse.}-- In this paper, we initiated a
study on phenomenological imprints of string axions which were located at plateau regions before they commence to oscillate. We found that for such axions, the resonance
instability can last without the disturbance of the cosmic expansion, because the delayed onset of the
oscillation makes the redshift of the resonant modes insignificant. This
instability takes place slightly below the Jeans scale, suggesting various
implications on the structure formation of ULA dark matter. The persistent
resonance instability leads to the emission of the detectable GWs
and the subsequent oscillon formation~\cite{KSU}. In
contrast to the GWs emitted during the reheating, the GWs at later
times, discussed in this paper, can be emitted in the directly detectable ranges.


{\it Acknowledgements.}-- We would like to thank A. Aoki and H. Tashiro
for fruitful discussions and especially N. Kitajima for the collaboration in our follow-up study in \cite{KSU}. Y.U. also thanks the organizers and 
participants of the workshop ``Post-inflationary string Cosmology,'' during which this work was completed. This work was in part supported by JSPS KAKENHI Grant
Numbers 17H02894, 17K18778 (J.S.), 16K17689 (Y.U.), and MEXT KAKENHI Grant Numbers
15H05895, 17H06359 (J.S.), 16H01095(Y.U.). Y.~U. is also supported by
Daiko foundation.

\end{document}